# Theoretical consideration of pits recording and etching processes in chalcogenide vitreous semiconductors

A.N. Morozovska*, S.A. Kostyukevych**

V. Lashkaryov Institute of Semiconductor Physics, NAS of Ukraine,

41, pr. Nauki, 03028 Kiev, Ukraine.

*morozo@mail.i.com.ua, **sekret@spie.org.ua.

We propose theoretical consideration and computer modeling of information pit recording and etching processes in chalcogenide vitreous semiconductors. We demonstrate how to record and develop information pits with the necessary shape and sizes in chalcogenide photoresists using gaussian laser beam and selective etching. It has been shown that photo-transformed region cross-section could be almost trapezoidal or parabolic depending on the photoresist material optical absorption, recording beam power, exposure, etchant selectivity and etching time. Namely, during the laser illumination and thermal heating caused by it, photosensitive material is the quasi-equilibrium microscopic mixture of the transformed and non-transformed phases with different optical absorption coefficients: temperature dependent near the absorption edge "transformed" coefficient $\alpha_e$ and almost independent coefficient $\alpha_0$. If $\alpha_e \leq \alpha_0$ after thermal heating, the photo-transformed region "bleaches" and the pit depth increases more rapidly under the following laser power increasing. If $\alpha_e > \alpha_0$, the photo-transformed region "darkens" and the pit depth increases sub-linearly or even saturates under the following laser power increasing. Thus, almost parabolic or flattened pits appear in the case $\alpha_e \geq \alpha_0$, whereas the pits with elongated tops appear in the case $\alpha_e << \alpha_0$. After illumination, the spatial distribution of photo-transformed material fraction was calculated using the Kolmogorov-Awrami equation. Analyzing obtained results, we derived a rather simple approximate analytical expression for the dependence of the photo-transformed region width and depth on the recording gaussian beam power, radius and exposure time. Then the selective etching process was simulated numerically. The obtained results quantitatively describes the characteristics of pits recorded by the gaussian laser beam in thin layers of $As_{40}S_{60}$ chalcogenide semiconductor. Our model open possibilities how to select the necessary recording procedure and etching conditions in order to obtain pits with the optimum shape and sizes.

## 1. INTRODUCTION

Laser recording of profiled microstructures are widely used in modern manufacturing of optical compact disks (CD) and originals of diffraction elements, synthesized holograms, etc [1]. Numerous experimental and theoretical investigations [2-8] devoted to the information recording and reading by optical methods has already been performed. Despite numerous of achievements, a great amount of different problems concerning recorded information quality and its density increase still remains unsolved. It was shown [8-12], that using the inorganic resists based on $As_{40}S_{60}$ chalcogenide glass allows to obtain the profiled structures with sizes much smaller than the diameter of the recording focused laser beam at the wavelengths $\lambda = (436 \div 532) nm$.

This paper is devoted to the theoretical consideration of the question how to record pits with the necessary depth profile in photosensitive materials by varying their activation energy, optical absorption, recording gaussian beam power and exposure as well as the etchant selectivity and etching time. We used the following assumption: the change of thin film structure and temperature caused by a laser beam is the main factor that determines the photo-thermo-transformed material amount. We assume that in the course of illumination photosensitive material is quasi-equilibrium microscopic mixture of the transformed and non-transformed phases with relative fractions $M$ and (1-$M$) respectively. These phases have different photo-chemical properties and the optical absorption coefficients $\alpha_e$ and $\alpha_0$ [6]. The approximate formula for the photo-thermo-transformed region that determines the profile parameters (width and height) of the pit has been obtained. It has been shown, that pit profile depending on photosensitive material parameters, recording beam power, exposure and etching conditions could be flat with rounded edges or parabolic (see Sch. 1).

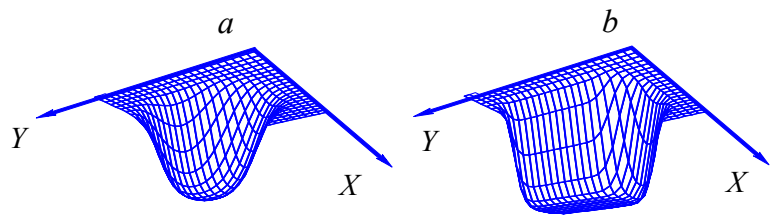

**Sch. 1**. The photo-thermo-transformed region shape, i.e. $M(r,z)$ spatial distribution calculated at sufficiently small (plot a) and high (plot b) beam exposures.

## 2. FUNDAMENTAL EQUATIONS

When the intensive focused laser beam illuminates the photosensitive layer, the optical absorption in the layer $0 < z < \ell$ is described by the Bouguer-Lambert-Beer law:

$$\frac{\partial I(x,y,z,t)}{\partial z} = -\alpha(x,y,z,t) I(x,y,z,t), \quad (1a)$$

$$z \geq 0, \quad 0 \leq t \leq t_H.$$

Hereinafter $I(x,y,0,t)$ is the intensity of the laser beam focused onto the surface $z=0$. In accordance with the adopted assumption [6], the light absorption coefficient $\alpha$ is as follows:

$$\alpha(x,y,z,t) = \alpha_0 (1 - M(x,y,z,t)) + \alpha_e \cdot M(x,y,z,t). \quad (1b)$$

It is obvious, that $\alpha$ depends on the exposing beam wavelength $\lambda$, intensity, temperature and exposure time $t_H$ owing to its dependence on $M$. The system (1) can be rewritten using the light transmission coefficient $\beta$, namely:

$$I(x,y,z,t) = \beta(M,z) \cdot I_0(x,y,t), \quad (2a)$$

$$\beta(M(x,y,z,t), z) = \exp\left[-\alpha_0 \cdot z - (\alpha_e - \alpha_0) \cdot \int_0^z M(x,y,\xi,t) d\xi\right]. \quad (2b)$$

Hereinafter $I_0(x,y,t) = I(x,y,0,t)$ is the incident laser beam intensity, $0 < M(x,y,z,t) < 1$ and $\alpha_0 < \alpha_e$. Let us consider the model situation when the axial-symmetrical gaussian laser beam moves along $Y$-axis at the distance $d$ with the constant velocity $\vartheta$. The full exposure $H(x,y,z)$ could be calculated directly from (11) and (2), namely:

$$H(x,y,z) = \int_0^{t_H} dt\, \beta(M(x,y,z,t), z) \cdot I_m \exp\left(-\frac{x^2 + (y - \vartheta t)^2}{\rho_0^2}\right). \quad (3)$$

Hereinafter $I_m$ is the recording beam maximum intensity, $\beta$ is the light transmission coefficient (2b), $\rho_0$ is the recording beam characteristic radius, $t_H$ is the exposure time. The integration in (3) could be performed exactly when $\beta = 1$, i.e. at $z=0$ [7]. If only $d/\rho_0 \gg 1$ and $\rho_0 < y < d - \rho_0$, $H(x,y,0)$ profile almost coincides with the intensity profile in $X$-direction. Thus, one obtains that

$$H(x,y,0) \approx \frac{\sqrt{\pi}\rho_0}{\vartheta} I_m \exp\left(-\frac{x^2}{\rho_0^2}\right). \quad (4)$$

One can obtain from (4) that the exposure spatial distribution is gaussian. We assume that the distribution of the photoresist temperature $T(x,y,z,t)$ is determined by the light absorption [3, 9]. Dissipation of the thermal flux during the laser pulse time $t_H$ could be neglected, if the characteristic penetration depth of the thermal flux $\chi_T \sim \sqrt{t_H}$ is higher than the film thickness $\ell$. In this case, the stationary temperature distribution exists [7], namely:

$$T(x,y,z,t) \approx T_0 + \Delta T(x,y,z,t), \quad \Delta T(x,y,z,t) = \frac{I(x,y,z,t)}{I_P}\frac{E_a}{k_B}. \quad (5)$$

$E_a$ is the activation energy. The function $I_P$ formally introduced in (5) has the meaning of some characteristic intensity, any increase above which can provide registration of sizable amounts of photo-transformed material in our recording conditions.

The fraction of photo-transformed material $M$ could be calculated using the Kolmogorov-Awrami equation [1, 3, 7]. Allowing for (2) and (5) $I(x,y,z,t)$ and $T(x,y,z,t)$ depend on $M(x,y,z,t)$, and so the fraction should be determined from the Kolmogorov-Awrami equation in a self-consistent manner, namely:

$$M(x,y,z,t) \approx 1 - \exp\left(-\int_0^{t_H} \frac{d\tau}{\tau_m} \cdot \frac{I_0(x,y,\tau)}{I_P}\beta(M(\tau),z)\exp\left(-\frac{1}{\frac{k_B T_0}{E_a} + \frac{I_0(x,y,\tau)}{I_P}\beta(M(\tau),z)}\right)\right). \quad (6)$$

The characteristic time $\tau_m$ depends not only on properties of recording medium but also on the recording light wavelength $\lambda$. One immediately obtains from (4) and (6) that the total fraction $M_0(x,z) \equiv M(x,y,z,t_H)$ of photo-thermo-transformed region cross-section could be estimated by the iteration method starting from the point $z=0$:

$$M_0(x,z=0,H_m) \approx 1 - \exp\left(-\frac{H(x,0)}{H_0}\exp\left(-\frac{E_a}{k_B T(H(x,0))}\right)\right)$$

$$H(x,0) = H_m \exp\left(-\frac{x^2}{\rho_0^2}\right), \quad T(H(x,0)) = T_0\left(1 + \frac{H(x,0)}{H_T}\right) \quad (7)$$

Hereinafter the characteristic exposures $H_0 = \tau_m I_P$, $H_T = (k_B T_0 / E_a) I_p t_H$ and the maximum exposure $H_m \approx (\sqrt{\pi}\rho_0 / \vartheta) I_m t_H$ are introduced. Characteristic exposure $H_T$ is inversely proportional to the activation energy. It is noteworthy, that at $H_T \ll H_0$ significant thermal heating occurs even at small exposures $H_m < H_0$, and as a result visible structural transformations ($M_0 \sim 0.5$) inside the exposed region start at $H_m < H_0$ (see the light-blue curve in Fig.1). In contrast to this, at $H_T \gg H_0$ no thermal heating occurs at exposures

$H_m < H_T$ and visible structural transformations inside the exposed region starts only at $H_m \gg H_0$ (see the violet curve in Fig.1). The intermediate case is realized at $H_T \sim H_0$, when at small exposures $H_m < H_p$ only weak photo-structural transformations take place, then at $H_m \geq H_p$ thermal heating occurs (see the dark-blue curve at $H_m/H_0 > H_P$ in Fig.1).

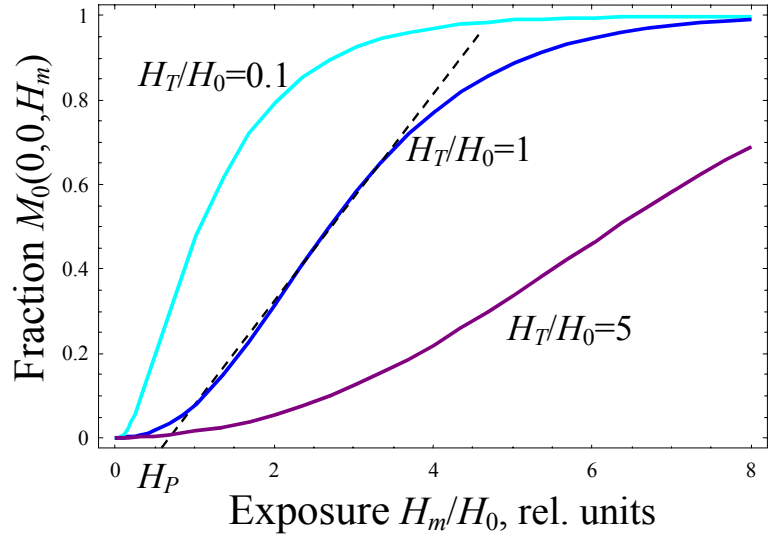

**Fig. 1**. Fraction $M_0(0,0,H_m)$ from (7) vs. the exposure $H_m/H_0$, at $E_a/k_B T_0 = 5$ and different $H_T/H_0$ ratios: 0.1 (light blue), 1 (dark-blue), 5 (violet).

Thus, for the majority of photosensitive materials the conditions $H_m < H_T$ and $H_T > H_0$ determine the cases of **photo-transformations** with a weak sub-linear or linear dependence of $M_0(0,0,H_m) \sim H_m$. This case is realized in laser lithography with wide homogeneous beams and holographic grating recording. Whereas the conditions $H_m \geq H_T$ and $H_T < H_0$ determine the cases of **photo-thermo-transformations** realized in CD recording by a focused laser beam.

Using (7) as a starting point, one obtains from (2b), (6) and (7) the expression for $M_0$ in the first approximation over the parameter $(\alpha_e - \alpha_0)/\alpha_0$:

$$M_0(x,y,z) \approx \exp(-\alpha_0 \cdot z - (\alpha_e - \alpha_0) \cdot M_0(x,y,0) \cdot z) \cdot M_0(x,y,0). \qquad (8)$$

As it follows from (8), if $\alpha_e \leq \alpha_0$ after the thermal heating, the exposed region "bleaches", and the pit depth increases more rapidly under the following laser power increasing. If $\alpha_e > \alpha_0$, the exposed region "darkens" and the pit depth increases sub-linearly or even saturates under the following laser power increasing. Thus, almost parabolic or flattened exposed region could appear in the case $\alpha_e \geq \alpha_0$, whereas the pits with elongated tops appear in the case $\alpha_e \ll \alpha_0$.

The experimental possibilities of photoresist etching in definite selective etchants is characterized by the etching rates: $\vartheta_n$ of non-exposed regions with $M \to 0$ and $\vartheta_e$ for the exposed ones with $M \to 1$. Similarly to the model (1b) used for absorption coefficient $\alpha$, the etching rate $\vartheta(x,z)$ in the point with the relative fraction $M_0(x, H_m, z)$ has the form:

$$\vartheta(x, H_m, z) = \vartheta_n (1 - M_0(x, z, H_m)) + \vartheta_e \cdot M_0(x, z, H_m). \tag{9}$$

After selective etching with time $t_E$, the exposed layer depth profile $z(x, y, t_E)$ can be determined using (9) as:

$$\frac{d\vec{r}(x, H_m, z, t)}{dt} = (\vartheta_n (1 - M_0(x, z, H_m)) + \vartheta_e \cdot M_0(x, z, H_m)) \cdot \vec{n}(\vec{r}, t)$$
$$z(x, y, 0) = 0, \quad \vec{n}(\vec{r}, 0) = \{0, 0, 1\}, \quad 0 \leq z \leq l_0, \quad 0 \leq t \leq t_E \tag{10}$$

Hereinafter $l_0$ is the initial thickness of the photoresist layer, $\vec{n}(x, y, z)$ is the etched region external normal. The approximate analytical expression for residual layer height profile can be obtained from (7)-(10) in the case of mainly normal-directed etching at $x^2/\rho_0^2 < 1$, when the equation (10) transforms into the scalar differential equation

$$\frac{dz(x,t)}{dt} \approx \vartheta_n + (\vartheta_e - \vartheta_n) M_0(x, 0, H_m) \exp(-(\alpha_0 + (\alpha_e - \alpha_0) \cdot M_0(x, 0, H_m)) \cdot z(x, t))$$
$$z(x, y, 0) = 0 \tag{11}$$

The equation (11) has the following analytical solution for the etched relief height profile:

$$z(x,t) = \frac{1}{\alpha(x, H_m)} \ln\left(\left(\frac{\vartheta_e}{\vartheta_n} - 1\right) M_0(x, 0, H_m)(\exp(\alpha(x, H_m)\vartheta_n t) - 1) + \exp(\alpha(x, H_m)\vartheta_n t)\right)$$
$$\tag{12}$$
$$\alpha(x, H_m) = \alpha_0 + (\alpha_e - \alpha_0) \cdot M_0(x, 0, H_m)$$

In the important for applications case of a highly selective positive etchant with $\vartheta_e/\vartheta_n \gg 1$ and for the exposed region with $\alpha(x, H_m)\vartheta_n t \leq 1$, $\alpha_e \approx \alpha_0$, the expression (12) describing the relief height profile can be approximated as:

$$z(x,t) \approx \vartheta_n t + (\vartheta_e - \vartheta_n) t \cdot M_0(x, 0, H_m). \tag{13}$$

The expression for $M_0(x,0,H_m)$ is given by (7). First, let us consider the region of photo-thermo-transformations with $0 \leq H_m \leq (1 \sim 10)H_T$, which is used for CD holographic recording and further development in a highly-selective positive etchant with $\vartheta_e/\vartheta_n >> 1$. In this case, the approximation (13) is valid. The dependence of the residual layer thickness on the exposure $H_m$ at the fixed etching time $t_E$ is presented in Fig. 2 for different $H_T/H_0$ ratios.

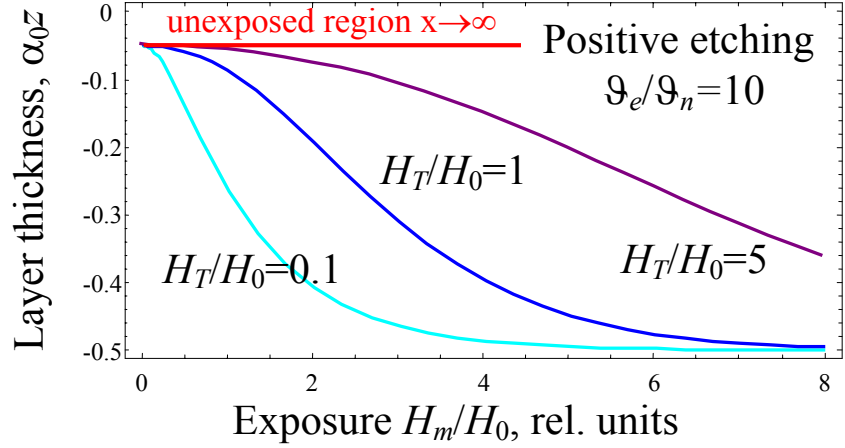

**Fig. 2**. Residual layer thickness (13) in $\alpha_0 z$ units vs. the exposure $H_m/H_0$ after positive selective etching ($t_E = 0.5\alpha_0/\vartheta_n$) at $E_a/k_B T_0 = 5$, $\alpha_e = \alpha_0$ and different $H_T/H_0$ ratios: 0.1 (light blue), 1 (dark-blue), 5 (violet).

It is clear from the figure that the smaller is $H_T/H_0$ ratio, the more significant is the contribution of thermal transformations and so residual layer thickness decreases more rapidly with the exposure increase.

Hereinafter, let us consider the region of photo-thermo-transformations with $H_T << H_0$. It is easy to obtain that at fixed exposure $H_m$ the residual layer thickness decreases with the etching time increase (see Fig. 3).

The residual layer thickness dependence over the exposure $H_m$ at fixed etching time $t_E$ is shown in the Fig. 4. Various cases can be realized for positive and negative etching, respectively.

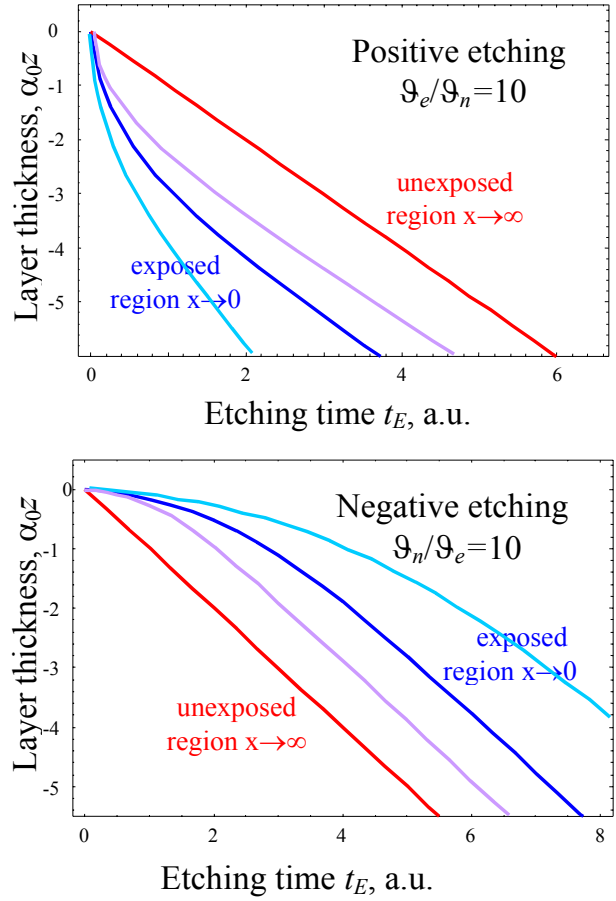

**Fig. 3**. Residual layer thickness (12) in $\alpha_0 z$ units via selective etching time measured in units $\alpha_0/\vartheta_n$ at fixed exposure $H_m/H_0 = 5$, $H_T \ll H_0$ and $\alpha_e = 0.5\alpha_0$ (light blue), $\alpha_e = \alpha_0$ (dark-blue), $\alpha_e = 2\alpha_0$ (violet).

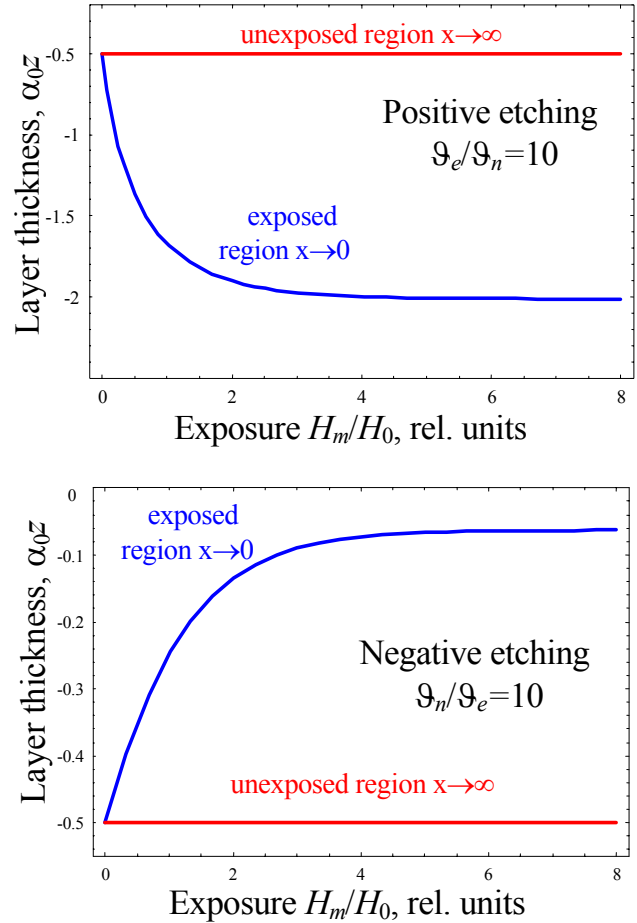

**Fig. 4**. Residual layer thickness (12) in $\alpha_0 z$ units vs. the exposure $H_m/H_0$ at fixed selective etching time $t_E = 0.5$ measured in units $\alpha_0/\vartheta_n$ and $\alpha_e = \alpha_0$, $H_T \ll H_0$ (compare with Fig. 2).

Photo-thermo-transformed region shapes at fixed exposure $H_m/H_0$ and increasing selective etching times $t_E$ are presented in Fig. 5. It is seen that the modulation depth increases with the etching time at the initial stage, then the etched profile "falls down" as a whole.

Photo-thermo-transformed region shapes at the fixed selective etching time $t_E$ and different exposures $H_m/H_0$ are presented in Fig. 6. It is seen that the modulation depth increase with the exposure, while the profile becomes more trapezoidal. Thus, almost parabolic or flattened pits could appear in the case $\alpha_e \geq \alpha_0$, whereas the pits with elongated tops appear in the case $\alpha_e \ll \alpha_0$.

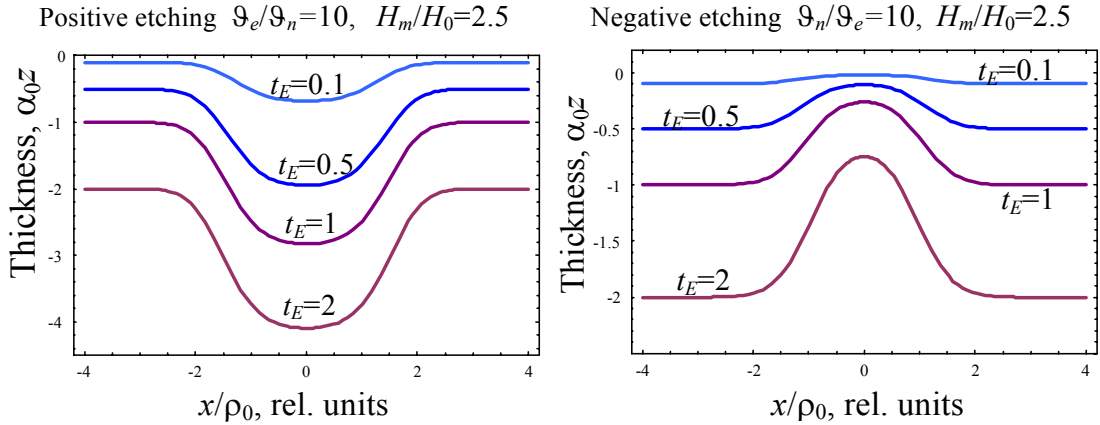

**Fig. 5**. Photo-thermo-transformed region shape (12) at the fixed exposure $H_m/H_0$ and increasing selective etching times $t_E$ measured in units $\alpha_0/\vartheta_n$, $\alpha_e = \alpha_0$, $H_T \ll H_0$.

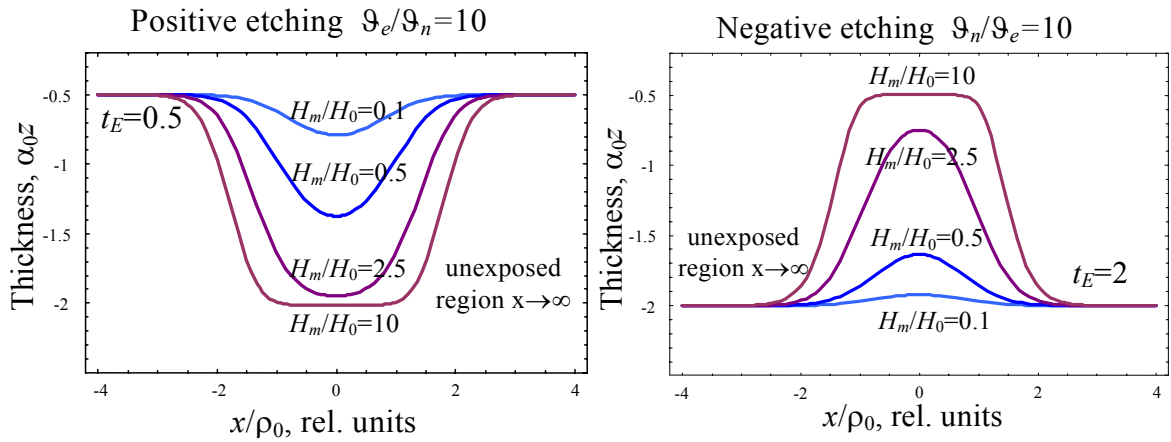

**Fig. 6**. Photo-thermo-transformed region shape (12) at the fixed selective etching time $t_E$ measured in units $\alpha_0/\vartheta_n$ and different exposures $H_m/H_0$, $\alpha_e = \alpha_0$, $H_T \ll H_0$.

## 3. THE CHARACTERISTICS OF RECORDED PITS

Finally, let us discuss the situation of information recording associated with the laser recording on a master disc possessing a thin chalcogenide layer of $As_{40}S_{60}$ composition [9-12]. In our recent experiments, the exposure of these thin layers was performed using focused laser light with the wavelength $\lambda = (436...488)\ nm$, optical absorption coefficient varied within the range $\alpha_0 \approx (3\cdot 10^5...3\cdot 10^3)\ cm^{-1}$ [4], [13]), beam diameter was close to $0.8\ \mu m$ and laser beam powers varied from 0.2 to 1 mW. Then the development of photoresist in positive selective etchant has been carried out. Under the positive etching, the exposed regions were completely removed, while the unexposed ones remained on the substrate. The etching rate of the exposed region is much smaller than that of the unexposed one. So, the profile of the photo-thermo-transformed region after positive selective etching could be determined from (12)-(13) in the first approximation. Below, we have presented results for $As_{40}S_{60}$ thin layers (see Figs 7, 8, 9).

The pit depth $z_p$ and halfwidth $a_p$ increase with the laser beam power $P$ (see Fig. 8). Starting from a definite laser power $P_P \sim H_P$ (see Fig. 1), thermal transformations play an important role in the structural transformations inside the exposed spot. We obtained that for the examined $As_{40}S_{60}$ layer $P_P \approx 0.1 mW$, $P_T \approx 0.7 mW$ ($H_T \approx P_T t$, $H_m \approx Pt$). As it follows from both our theoretical calculations (13) and experiments, the pit height $z_p$ increases linearly with $P$ in the region $P_p < P < P_T$, then saturates under further $P$ growth. The pit halfwidth $a_p$ gradually increases with the laser beam power $P$.

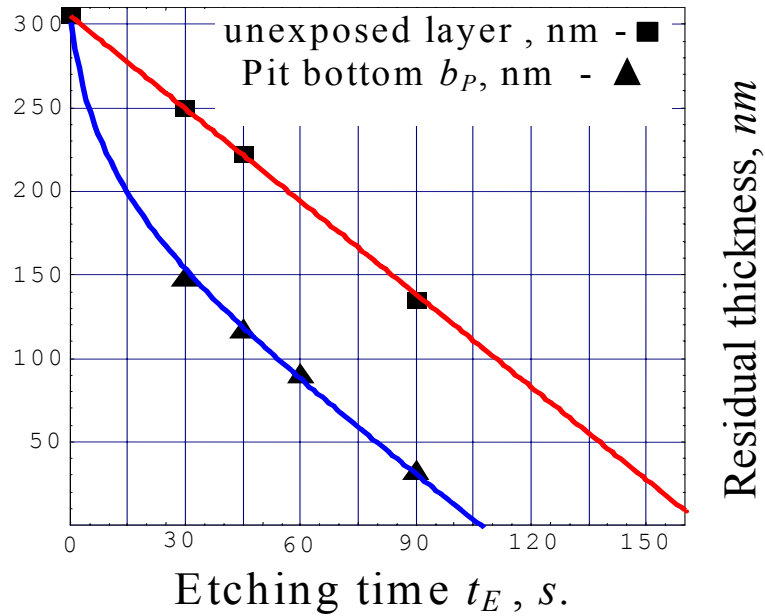

**Fig. 7**. Dependence of the exposed pit bottom (triangles, nm) and unexposed disk surface (squares, nm) vs. the etching time $t_E$ for $As_{40}S_{60}$ layer ($\lambda$=457 nm, $h_0$=305 nm, positive etching, laser power $P$=0.46 $mW$). Solid lines are our fitting (12) at $\vartheta_e = 18 nm/s$, $\vartheta_n = 1.85 nm/s$, $H_m/H_0 = 5$ and $\alpha_0 \approx 10^5 \, cm^{-1}$, $\alpha_e \approx 2\alpha_0$.

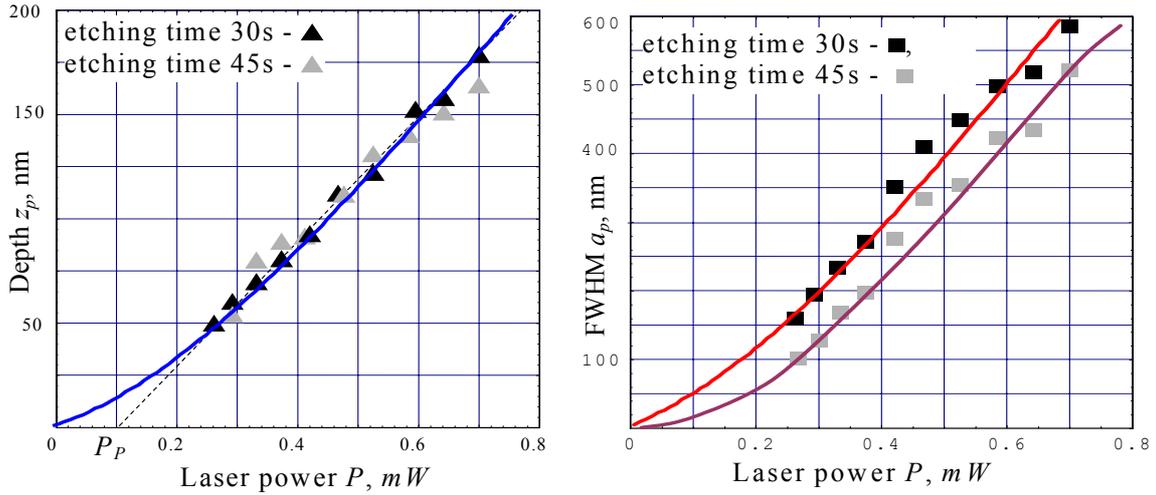

**Fig. 8**. Pit depth and halfwidth FWHM vs. the gaussian beam power for $As_{40}S_{60}$ layer ($\lambda = 457 nm$, $h_0 = 305 nm$, etching time 30s (black triangles and squares) and 45s (grey triangles and squares)). Solid curves are our fitting (13) and (10) at $E_a/k_B T_0 = 1.9$, $P_T = 0.7 mW$, $\vartheta_e = 18 nm/s$, $\vartheta_n = 1.85 nm/s$ (other parameters are the same as in Fig. 7).

Using atomic force microscopy (AFM), we examined the obtained microstructures. At the chosen beam power region and exposure time, the cross-section of the photo-thermo-transformed region is close to parabolic at small beam powers $P \leq 0.5\ mW$ (see Fig. 9), whereas it flattens at higher laser powers [10-12]. The small discrepancy between the pit cross-section obtained using AFM (black curve) and the calculated one (blue curve) could be regarded to the radial etching components neglected in our model (12). It is clear that the proposed model describes adequately the available experimental data [8].

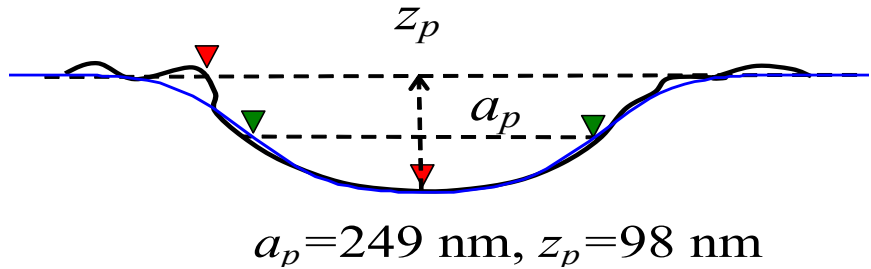

$a_p$=249 nm, $z_p$=98 nm

**Fig. 9**. Cross-section of the cavity developed using positive etching of $As_{40}S_{60}$ layer ($\lambda = 457 nm$, etching time 45 s, beam power 0.46 $mW$). The blue line is our fitting calculated in accordance with expression (12) at $H_m/H_0 = 1$, $\alpha_e = \alpha_0$, $H_T/H_0 = 1$.

## CONCLUSION

- The rather simple analytical expressions (12)-(13) for the pit height profile have been derived. In the most important for CD recording case of photo-thermo-transformations pits halfwidth and height are determined by the beam power, exposure, radius $\rho_0$, optical absorption coefficients $\alpha_{e,0}$, etching time and rate known for the recording material. The proposed model quantitatively describes the characteristics of pits recorded by a gaussian laser beam in the thin chalcogenide layers of $As_{40}S_{60}$ composition [7-8].

- Evolved approach allows one to select the necessary recording conditions in order to obtain the pits with the optimum shape after positive selective etching of the chalcogenide photoresist, which is useful for applications.

**Acknowledgements.** Authors are grateful to Dr A.V. Stronsky, PhD P.E. Shepeliavyi and PhD A.A. Kudryavtsev for useful discussions of our model.